\documentstyle{article}
\input epsf

\begin{document}
\parbox{13 cm}
{
\begin{flushleft}
\vspace* {1.2 cm}
{\Large\bf
{
Antisymmetric multi-partite quantum
states and their applications
}
}\\
\vskip 1truecm
{\large\bf
{
I. Jex$^1$), G. Alber$^2$), S. M. Barnett$^3$) and A. Delgado$^4$)
}
}\\
\vskip5truemm
{
$^1$) Department of Physics, FNSPE, CTU Prague, B\v rehov\'a 7, 115 19 Prague, Czech Republic\\
$^2$) Institut f\"ur Angewandte Physik,
Technische Universit\"at Darmstadt, D-64289 Darmstadt,Germany\\
$^3$) Department of Physics and Applied Physics, University of Strathclyde, Glasgow Q4 ONG, UK \\
$^4$) Department of Physics and Astronomy, University of New Mexico, 87131 Albuquerque, USA
}
\end{flushleft}
}
\vskip 0.5truecm
\noindent
{\bf Abstract:\\}
{\noindent
Entanglement
is a powerful resource
for processing quantum information.
In this context pure, maximally entangled states have received considerable attention.
In the case of bipartite qubit-systems
the four orthonormal Bell-states are of this type.
One of these Bell states, the singlet Bell-state,
has the additional property of
being antisymmetric with respect to particle exchange.
In this contribution we discuss possible generalizations
of this antisymmetric Bell-state to cases with more than two particles and with single-particle
Hilbert spaces involving
more than two dimensions.
We review basic properties of these totally antisymmetric states. 
Among possible applications of this class of states we analyze 
a new quantum key sharing protocol and methods for comparing
quantum states.}
\vskip0.1cm
\noindent
PACS: 03.67.-a, 03.65.Ta
\section{Introduction}

By now,
quantum theory has become a well established part of modern physics. We have
become accustomed to its results even if some of the concepts involved appear strange from the point of view
of classical physics. 
However, as long as these peculiarities are restricted to the microscopic domain it is not so difficult for
us to get used to them. 
During the last decade there have been various successful attempts to push characteristic quantum phenomena into
the macroscopic domain and to exploit these very phenomena for practical purposes.
These attempts may be viewed as first steps of a newly emerging  quantum technology.
Thus,
it was possible
to propose new, efficient quantum algorithms, to develop methods for the transfer of quantum states and of secret keys,
and to invent new quantum error correction methods which suppress decoherence.
Quite a number of these new effects rely on the use of states whose correlations are incompatible with local realistic
theories. The singlet state of two distinguishable spin-1/2 particles is a prominent example which has
been studied extensively in the past.
The main purpose of the subsequent contribution
is to point out
several possible applications of generalizations of this singlet state to cases which involve
more than two distinguishable quantum systems of arbitrary but finite dimensions.

\section{Definition and basic properties of totally antisymmetric quantum states}

Totally antisymmetric quantum states are natural generalizations of the singlet state to many-particle quantum systems.
In atomic and molecular physics, for example, they have already been playing an important role
as Slater-determinant states. These are defined by the relation 
\begin{equation}
\vert A_N\rangle = \frac{1}{\sqrt{N!}} \sum_\pi (-1)^{sgn (\pi )} \vert \pi_1\rangle ...\vert\pi_N\rangle  \label{defA}
\label{definition}
\end{equation}
with $\{|i_1\rangle ... \vert i_N\rangle; i_1,...,i_N = 0,...,d-1\}$ denoting an orthonormal basis of the 
Hilbert space of $N$ $d$-dimensional quantum systems. The sum appearing in Eq.(\ref{definition})
runs over all possible permutations $\pi$ of the $N$ elementary quantum systems considered.
Due to basic properties of determinants
this state exists only in cases in which 
the number of particles $N$  equals the dimension of the
one-particle Hilbert spaces $d$ involved. Thus,
in the simple case of three qutrits, for example, the totally antisymmetric quantum state is given by
\begin{equation}
\vert A_3\rangle = \frac{1}{\sqrt{6}} \left\{\vert 0\rangle\vert 1\rangle\vert 2\rangle + \vert 1\rangle\vert 2\rangle\vert 0\rangle +
\vert 2\rangle\vert 0\rangle\vert 1\rangle  - \vert 0\rangle\vert 2\rangle\vert 1\rangle - \vert 1\rangle\vert 0\rangle\vert 2\rangle - \vert 2\rangle\vert 1\rangle\vert 0\rangle\right\} . \label{A3}
\end{equation}
Let us summarize briefly some of the  most important properties of these totally antisymmetric states:
\begin{enumerate}
\item
They are invariant under local unitary transformations of the form $U\otimes U\otimes ...\otimes U$, i.e.
\begin{equation}
U\otimes U\otimes ...\otimes U \vert\psi\rangle \langle \psi\vert
U^{\dagger}\otimes U^{\dagger}\otimes ...\otimes U^{\dagger} 
= \vert\psi\rangle \langle \psi\vert.
\label{invpr}
\end{equation}
\item
Simultaneous measurements of all particles 
in a commonly chosen measurement basis
result in perfect correlations, i.e.
\begin{equation}
P(i_1,\dots,i_N)\equiv \mid \langle \pi_1| ...\langle \pi_N|\psi\rangle \mid^2
\frac{1}{N!}|\varepsilon_{i_1,\dots,i_N}|^2  \label{correl}
\end{equation}
with $\varepsilon_{i_1,\dots,i_N}$ denoting the totally antisymmetric tensor which is non-zero only
if all its indices are different.
\item
They
can be generated in an iterative manner by a sequence of generalized XOR-gates and discrete Fourier transforms. 
The three-particle state $\vert A_3\rangle$, for example, can be prepared from the antisymmetric two-particle state 
$\frac{1}{\sqrt{2}} (\vert 2\rangle_1 \vert 1\rangle_2 - \vert 1\rangle_1 \vert 2\rangle_2 )$ by
$$
\vert A_3\rangle = GR_{31}~ GR_{32}~F_3~\vert 0\rangle_3  
\frac{1}{\sqrt{2}} (\vert 2\rangle_1 \vert 1\rangle_2 - \vert 1\rangle_1 \vert 2\rangle_2 ).
$$
Thereby,
$F_3$ denotes the discrete Fourier transformation applied to  the third particle
and $GR_{ij}$ represents a generalized XOR-operation applied to particles $i$ and $j$.
Applied to the first and second particle, for example,
this latter operation is defined by
\begin{equation}
GR_{12}~\vert i\rangle_1 \vert j\rangle_2 = \vert i\rangle_1 \vert i\ominus j~\rangle_2  
\end{equation}
with $\ominus$ denoting subtraction $mod (d)$.
This construction can be generalized in a
straightforward way to 
more than three particles.
\item
In the case of $N$ particles
the reduced density matrix $\hat\rho_i$ of subsystem $i$ is given by
\begin{equation}
\hat\rho_i = \frac{1}{d}\sum\limits^{d-1}_{j=0} \vert j\rangle\langle j\vert ,
\end{equation}
i.e., the single-particle reduced density matrix describes a completely depolarized state.
Projection of one of the particles onto a particular state, say $\vert j\rangle\langle j\vert$, leaves the rest of the 
system in the pure antisymmetric state which involves all one-particle states except state $|j\rangle$, i.e.
\begin{equation}
\vert \overline{A}_{N-1}\rangle = \frac{1}{\sqrt{(N-1)!}} \sum_\pi (-1)^{sgn (\pi )} \vert \pi_1 \rangle ...\vert \pi_{N-1} \rangle.
\end{equation}
The index of correlation \cite{Steve} between a particular particle and
the remaining part of the system is given by
\begin{equation}
I_{i-r} = S_i + S_r - S = 2 S_i = 2 {\rm log~(d)} . \label{IC}
\end{equation} 
with  the von-Neumann entropy of particle $i$ being given by $S_i = - Tr{\hat\rho_i \ln\hat\rho_i}$
and with
$S_{i-r}$ denoting the
von-Neumann entropy of the remaining part. The entropy $S$ of the whole system equals zero as it is in a pure state. 
\end{enumerate}
As exemplified in the subsequent sections
totally antisymmetric quantum states can be used for many tasks which are of interest for quantum communication.

\section{A quantum  mechanical key sharing protocol}

The secret
distribution of a classical key is one of the main aims of quantum cryptography. Known secure protocols of
bipartite key distribution are either based on non-orthogonal two dimensional quantum states \cite{bennet} or 
on entangled states \cite{ekert}. These protocols enable two parties to share a common, secret classical key.
Recently, several more general situations have been discussed. One of them involves
the distribution of a
classical key between several parties in
such a way that a subset of the parties has access to the key only if they share the information available.
Various
multi-partite key sharing protocols of this kind have been proposed which are either based on the use of GHZ-states \cite{hbb} 
or on the use of pairs of singlet states \cite{karlsson}. 

Here we discuss an alternative
multi-partite
quantum key sharing
protocol which is based on anti-symmetric states of qudit
systems. (A qudit system is a $d$ dimensional elementary quantum system.)
This protocol enables one to generate, to split and to distribute 
a classical d-ary key securely.
We demonstrate the basic principles of this protocol for quantum key sharing
in
the simplest nontrivial case of three three-dimensional quantum systems.
In this case we base our quantum protocol on the totally antisymmetric state $\vert A_3\rangle$ defined by
Eq. (\ref{A3}). 
For this key sharing protocol two basic properties  of totally antisymmetric states are important. Firstly,
all outcomes of simultaneous measurements performed by the participants in identical bases must be different and
secondly,
the unitary invariance of A-states guarantees that this is also true for any commonly chosen basis.

Let us consider three parties (Alice, Bob
and Charlie). Each of them is 
endowed with a common set ${\cal U}$
of unitary transformations. The protocol runs as follows:

\begin{itemize}
\item Alice prepares three qutrits in the anti-symmetric state 
$\vert A_3\rangle$. She applies a unitary
transformation ($\in {\cal U}$) on qutrit one,
measures this qutrit and keeps her
choice of the unitary transformation and the measurement result secret.
This transformation with the subsequent measurement changes the correlations in the anti-symmetric
state.

\item Alice sends qutrit two to Bob and qutrit three to Charlie.

\item In order to recover the original state and the correlations of
the measurements, Bob (Charlie) also chooses a unitary transformation
($\in {\cal U}$) randomly  and applies it onto his qutrit.
Afterwards Bob (Charlie)
measures his qutrit.
Alice keeps
her choice  secret. 

\item Bob (Charlie) transmits his choice of
transformation to Alice but keeps the measurement outcome secret. If all
three unitary transformations coincide, Alice declares the outcomes
of the measurements to be a valid part of the key. In this case,
Bob and Charlie can deduce Alice's result if they share the
outcomes of their measurements.

\item In order to study the security of the key generated by
this protocol Alice requests from Bob and Charlie a subset of the
outcomes of their measurements. 
\end{itemize}

\section{Security of the quantum key sharing protocol}

As a general investigation of security is beyond the scope of this contribution,
we restrict our subsequent discussion to a cut-and-resend attack which does not involve
coherent measurements.
In such an attack an external or internal eavesdropper could try to obtain
information about the key by
attaching an ancilla state to the three qutrits. Subsequently, measurement of
the ancilla could reveal information about the outcomes of
measurements performed on the qutrits.

The most general state of a system composed of qutrits
and an ancilla is given by
\begin{equation}
|E\rangle\equiv\sum_{i_1,i_2,i_3=0}^{2}|i_1\rangle |i_2\rangle |i_3\rangle\otimes
|E_{i_1,i_2,i_3}\rangle . 
\end{equation}
Thereby, the ancilla system is described by the states
$|E_{i_1,i_2,i_3}\rangle$. These states need not be mutually
orthogonal but they obey a normalization condition, namely $\langle E | E\rangle =1$.
If the eavesdropper wants to
remain undetected he must design the state $|E\rangle$ in such a
way that the probabilities $P(i_1,i_2,i_3)$ remain unchanged. This
imposes a set of constraints onto the states
$|E_{i_1,i_2,i_3}\rangle$. If we choose ${\cal U}=\{{\bf 1},F\}$
with $F$ denoting the discrete Fourier transform these constraints are given by 
the equations
\begin{eqnarray}
|E_{012}\rangle+|E_{021}\rangle+
|E_{120}\rangle+|E_{102}\rangle+
|E_{201}\rangle+|E_{210}\rangle = 0, & &      \nonumber                  \\
|x|^2(|E_{012}\rangle+|E_{021}\rangle)+
x^*(|E_{102}\rangle+|E_{120}\rangle)+
x(|E_{210}\rangle+|E_{201}\rangle) = 0, & & \nonumber                       \\
x^*(|E_{012}\rangle+|E_{210}\rangle)+
|x|^2(|E_{102}\rangle+|E_{201}\rangle)+
x(|E_{120}\rangle+|E_{021}\rangle)= 0, & &  \nonumber                     \\
x(|E_{012}\rangle+|E_{102}\rangle)+
x^*(|E_{201}\rangle+|E_{021}\rangle)+
|x|^2(|E_{210}\rangle+|E_{120}\rangle)= 0, & & \nonumber                       \\
|x|^2(|E_{012}\rangle+|E_{021}\rangle)+
x(|E_{102}\rangle+|E_{120}\rangle)+
x^*(|E_{210}\rangle+|E_{201}\rangle)= 0, & & \nonumber                      \\
x(|E_{012}\rangle+|E_{210}\rangle)+
|x|^2(|E_{102}\rangle+|E_{201}\rangle)+
x^*(|E_{120}\rangle+|E_{021}\rangle)=0,& & \nonumber                       \\
x^*(|E_{012}\rangle+|E_{102}\rangle)+
x(|E_{201}\rangle+|E_{021}\rangle)+
|x|^2(|E_{210}\rangle+|E_{120}\rangle) = 0 & & \nonumber             
\end{eqnarray}
with $x\equiv e^{-i\frac{2\pi}{3}}$. The unique solution of this set of equations is
given by
\begin{equation}
|E_{i_1,i_2,i_3}\rangle=\varepsilon_{i_1,i_2,i_3}|R\rangle.
\end{equation}
This result implies that, provided the eavesdropper wants to remain undetected, the state
$|E\rangle$ has to have the form
\begin{equation}
|E\rangle=\left(\frac{1}{3!}\sum_{i_1,i_2,i_3=0}^2
\varepsilon_{i_1,i_2,i_3}|i_1\rangle |i_2\rangle |i_3\rangle
\right) |R\rangle. 
\end{equation}
Thus,
the state of the ancilla factorizes from the qutrit-system so that the
eavesdropper cannot obtain any information about the key.
If the eavesdropper wants to retrieve information about the
key he must perturb the state in such a way that the correlations of the
outcomes of the measurements are changed.

\section{A protocol for quantum state sharing}

Totally antisymmetric states
are also well suited for 
distributing $d$-dimensional quantum states between $N=d$ parties. The task of quantum state
sharing to be realized may be viewed as
a generalization of the well-known bipartite entanglement-assisted teleportation protocol.
The aim of the protocol is to send the state $\vert\chi\rangle$ 
from a source to a particular receiver. However, due to security reasons
it should be possible to reconstruct this state  only
if all participants cooperate. Thus,
reconstruction of the state $|\chi\rangle$ by the 
receiver should be possible only if at least one additional mediator communicates additional classical
information properly.
In the simplest  case of three parties, i.e. $N=d=3$, the protocol implementing this task is
characterized by the following identity
\begin{equation}
|\chi \rangle _{1} |A_3\rangle _{234}\equiv 
\sum_{l,\rho
=0}^{2}\frac{1 }{3}|\Psi _{l,\rho }\rangle _{12}\,
\sum_{k=0}^{2}\frac{1}{\sqrt{3}} e{}^{i\frac{2\pi }{3}k\rho
}F_{3}^{-1}|k\rangle_{3}\,U(l,\rho ,k)|\chi \rangle _{4}.
\label{help1}
\end{equation}
This identity involves four particles, namely particle one which carries the quantum state $|\chi\rangle$
and particles two, three and four which are distributed between the three parties involved in the protocol. 
The orthonormal states $|\Psi _{l,\rho }\rangle_{12}$  are defined by
\begin{equation}
|\psi _{l ,\rho }\rangle_{12} = 
\frac{1}{\sqrt{3}}\sum
\limits_{k=0}^{2}{}e^{i\frac{2\pi }{3}l k}|k\rangle _{1} 
|k\ominus \rho \rangle _{2}~~{\rm mod~3} . \label{gbell}
\end{equation}
These orthonormal states generalize the  Bell basis to the case of two
qutrits. The unitary transformation $U(l,\rho,k)$ is given by
\begin{equation}
U(l,\rho ,k)|m\rangle_{4}\equiv e{}^{-i\frac{2\pi }{3}
lm}\sum_{q,r=1}^{3}{}e^{i\frac{2\pi }{3}kq}\varepsilon_{m-\rho
,q,r}|r\rangle _{4}.
\end{equation}
$F^{-1}$ denotes 
the inverse discrete Fourier transform.

The identity of Eq.(\ref{help1}) suggests the following protocol for quantum state sharing:
The
sender obtains particle two of the totally antisymmetric quantum 
state. Particles three and four are sent to the other two parties.
The sender who is now holding particles one and two performs a maximal quantum test on these two
particles by projecting onto 
the orthonormal basis of generalized Bell states (\ref{gbell}). As a consequence he obtains
two measurement results, say  $l$ and $\rho$, which specify the Bell state particles one and two
have been projected onto. Now, one of the other parties applies
a discrete Fourier transformation onto particle three and performs a maximal quantum test on this particle.
The result of this measurement yields the label of the quantum state particle three has been projected onto, say $k$.
The three classical labels, namely $(l,\rho,k)$ are communicated to the receiver. Only after having
received this combined classical information from the other two parties is the receiver
able to apply the proper inverse transformation, namely $U^{\dagger}(l,\rho,k)$, onto his particle
which enables him to recover
the original quantum state $\vert\chi\rangle$. 

\section{Comparison of two quantum states I}
Quantum state identification and state comparison constitute two other 
interesting applications of totally antisymmetric quantum
states \cite{comp}. Thereby one wants to answer the basic question
whether {\it two given  quantum states are identical or different}.
The simplest version of this problem can be illustrated in the case of two qubits. 
We shall comment on two separate
cases, namely on the case of two unknown and on the case of two known pure states. 

Let us first assume that we are given two completely unknown pure quantum states and that we want to decide with maximum probability
whether these states are identical or different. 
In the case of two unknown states, say  $\vert\psi\rangle$ and $\vert\phi\rangle$,
we cannot give an affirmative answer to the question
whether {\it these two states are the same}. We can only determine whether theses states are different or whether
the answer is inconclusive. The 
fact that a positive answer to this question cannot be obtained 
can be demonstrated in several ways.
The most straightforward argument relies on continuity. For any pair of different states the affirmative answer should
yield a zero result even in cases in which these states are only infinitessimally different. As a consequence the probability for a non-zero
result would have to
be discontinuous. This contradicts the fact that quantum mechanical probabilities are continuous functions
of projection operators.

In view of this impossibility the natural question arises
how to proceed in order to obtain at least a negative and an inconclusive answer. 
The product state of two qubits $\vert\psi\rangle \vert\phi\rangle$ can be decomposed uniquely into the symmetric 
states
$\vert 0\rangle\vert 0\rangle ,\vert 1\rangle\vert 1\rangle ,
 (\vert 0\rangle\vert 1\rangle  
+ \vert 1\rangle\vert 0\rangle)$ and into the antisymmetric state $(\vert 1\rangle\vert 0\rangle  
- \vert 0\rangle\vert 1\rangle)$. If we find a non-zero projection onto the antisymmetric state, the two 
states cannot be identical. If the measurement yields an overlap with one of the symmetric states the answer
is inconclusive. What can we say about the relative frequency of these two possible outcomes?
The overlap between the decomposition components is 
given by
\begin{equation}
P_s - P_a = \vert\langle\psi\vert\phi\rangle\vert^2 \geq 0 ,
\end{equation}
where $P_s = 1 - P_a$ and $P_a = \vert (\langle 1\vert\langle 0\vert - \langle 0\vert\langle 1\vert)\vert \psi\rangle\vert\phi\rangle\vert^2/2$ is the overlap between the tested product state 
$\vert \psi\rangle\vert\phi\rangle$ and the antisymmetric state. Thus,
the measurement will show the inconclusive result (projection onto the symmetric subspace) more often than a 
negative one. 

A realization of this state comparison using passive optical elements
(detection in the Bell basis) seems feasible.  We have to distinguish in a reliable
way the presence of the antisymmetric state from any element of the symmetric
subspace. For this purpose
also a simple coincidence measurement could be used. The two states can be sent into a multiport, for example,
and at the output
the coincidences can be detected. Only if both states are identical certain coincidences are absent.

Procedures which are applicable to more than two copies require a more detailed study of the group structure of the 
corresponding state spaces. If the number of copies equals the dimension of the one-particle Hilbert spaces, i.e.
$N = d$, then a comparison is simple as a totally antisymmetric state $\vert A_N\rangle$ 
exists. Otherwise we have to use projections onto combinations of the ''most antisymmetric'' states available.
Let us consider the simple example of $N=2$ and  $d > 2$. The two-particle Hilbert space can be decomposed
into two subspaces, namely a symmetric one, spanned by the vectors
$\vert i\rangle\vert i\rangle$ and $(\vert i\rangle\vert j\rangle + \vert j\rangle\vert i\rangle )$, and an antisymmetric one, spanned by the states
$(\vert i\rangle\vert j\rangle - \vert j\rangle\vert i\rangle)$ with $i,j = 0,...,d-1$. Successful projection onto the latter state indicates that the two quantum states are
different. Another simple case arises if $N=3$ and $d=2$.
The eight dimensional three-particle Hilbert space can be decomposed into two subspaces spanned by
the states
$\vert  1\rangle\vert 1\rangle\vert 1\rangle ,
\vert  0\rangle\vert 0\rangle\vert 0\rangle ,
(\vert 1\rangle\vert 1\rangle\vert 0\rangle + \vert 1\rangle\vert 0\rangle\vert 1\rangle +
\vert 0\rangle\vert 1\rangle\vert 1\rangle), 
(\vert 1\rangle\vert 0\rangle\vert 0\rangle + \vert 0\rangle\vert 0\rangle\vert 1\rangle +
\vert 0\rangle\vert 1\rangle\vert 0\rangle)$ and by the states
$(2\vert 1\rangle\vert 1\rangle\vert 0\rangle - \vert 1\rangle\vert 0\rangle\vert 1\rangle - \vert 0\rangle\vert 1\rangle\vert 1\rangle ), (2\vert 0\rangle\vert 0\rangle\vert 1\rangle - 
\vert 0\rangle\vert 1\rangle\vert 0\rangle - \vert 1\rangle\vert 0\rangle\vert 0\rangle ), 
(\vert 1\rangle\vert 0\rangle\vert 1\rangle - \vert 0\rangle\vert 1\rangle\vert 1\rangle ), (\vert 0\rangle\vert  1\rangle\vert 0\rangle - \vert 1\rangle\vert 0\rangle\vert 0\rangle )$. The latter four dimensional
subspace can be used
to decide whether three two-level states are different.

\section{Comparison of two quantum states II}

Let us now assume that two qubits are each prepared in one of the known 
states 
\begin{equation}
\vert\psi_{1,2}\rangle = \cos\theta\vert +\rangle \pm \sin\theta \vert - \rangle .
\end{equation}
The problem of comparing these two states can be solved either by the strategy of minimum probability of error
or by the strategy of unambiguous state identification (for a review see Ref.\cite{Tony} and references therein).
In the first case the minimum error with which both states can be distinguished is given by
\begin{equation}
P^{comp}_e = \frac{1}{2}\cos^2 (2\theta ).
\end{equation}
In the second case the minimum probability of obtaining an inconclusive answer is given by
\begin{equation}
P^{comp}_? = \cos (2\theta ) [2 - \cos (2\theta )].
\end{equation}
The question is whether these two strategies are optimal.
Indeed, the minimum error strategy is optimal \cite{comp}.
In the case of unambiguous state identification strategy we can do 
better. In this latter case the optimum strategy is the following:
First we use the Bell state decomposition
\begin{eqnarray}
\vert\psi_{i}\rangle\vert\psi_j\rangle  &=&  \cos^2\theta\vert +\rangle \vert +\rangle + 
(-1)^{i+j} \sin^2\theta\vert -\rangle \vert -\rangle + \\ \nonumber
& & (-1)^i \cos\theta\sin\theta[-\delta_{ij} 
(\vert +\rangle \vert -\rangle + \vert -\rangle \vert +\rangle) + \\ \nonumber 
& & (1-\delta_{ij}) (\vert +\rangle \vert -\rangle - \vert -\rangle \vert +\rangle)].
\end{eqnarray}
If
we project successfully onto the antisymmetric state 
$(\vert +\rangle \vert -\rangle - \vert -\rangle \vert +\rangle)$, the two states are different. If we 
project onto the symmetric state $(\vert +\rangle \vert -\rangle + \vert -\rangle \vert +\rangle)$, both states
have to be identical.
If
the state is found neither in the symmetric nor in the antisymmetric subspace, it is in one of the two possible
states
\begin{equation}
\vert\Phi_{\pm}\rangle = \frac{\cos^2\theta \vert +\rangle\vert +\rangle \pm \sin^2\theta \vert -\rangle\vert -\rangle}
{\sqrt{1 - \frac{1}{2}\sin^22\theta}}
\end{equation}
which can be discriminated unambiguously. Thus, the overall probability for an inconclusive result reads
\begin{equation}
P^{comp1}_? = (1 -\frac{1}{2} \sin^2 2\theta )\vert\langle\Phi_+\vert\Phi_-\rangle\vert = \cos 2\theta 
\end{equation}
and
\begin{equation}
P^{comp1}_{?} < P^{comp}_? .
\end{equation}
These simple considerations 
illustrate that the unambiguous method of state comparison is not the optimal one. It can be shown, however, that the two-step
method proposed is the optimal one. The interesting aspect of our analysis is that the unambiguous state discrimination
may be viewed as a two-step state comparison.
First we find out whether the two states are identical or not and afterwards
we determine the label.

\section{Conclusions}

We have demonstrated that totally antisymmetric quantum states are useful for various tasks in quantum
information processing.
Their special features are particularly useful for implementing 
multi-partite key-sharing and quantum state sharing  protocols and for comparing quantum states. 
All the applications discussed here rely on the high symmetry and the peculiar correlation properties of these quantum
states. 
It is expected that
the future development of multi-partite protocols for quantum information processing
will stimulate many more interesting applications
of totally antisymmetric quantum states.
\vskip3truemm
\noindent{\bf Acknowledgments}

This work was supported by the European IST-1999-13021 QUIBITS, DLR (CZE00/023) and GA\v CR (202/01/0318).


\begin{thebibliography}{1}
\bibitem{Steve} {S. M. Barnett and S. J. D. Phoenix} {Phys. Rev. A} {\bf 40} (1989) 2404
\bibitem{bennet} {Ch. H. Bennett and G. Brassard}, {Int. Conf. Computers,
Systems \& Signal Processing}, Bangalore, India, (1984) 175
\bibitem{ekert} {A. Ekert}, {Phys. Rev. Lett.} {\bf 67} (1991) 661
\bibitem{hbb} {M. Hillery, V. Bu\v{z}ek, and A. Berthiaume}, {Phys. Rev. A}
{\bf 59} (1999) 1829
\bibitem{karlsson} {A. Karlsson, M. Koashi and N. Imoto}, {Phys. Rev. A}
{\bf 59} (1999) 162
\bibitem{Tony} {A. Chefles}, {Contemp. Physics} {\bf 41} (2000) 401
\bibitem{comp} {S. M. Barnett, A. Chefles, I. Jex}, quant-phys/0202087 
\end{thebibliography}
\end{document}